\documentstyle[12pt,epsfig]{article}

\begin{document}

\title{HARD GAMMAS FROM LOW-ENERGY HEAVY-ION COLLISIONS}

\author{E. B\v{e}t\'{a}k\\
Institute of Physics, Slovak Academy of Sciences,
84228 Bratislava, Slovakia\\
{\em e-mail} betak@savba.sk}

\date{\em Presented at Triangle Meeting --- School and Workshop
on Heavy-Ion Collisions, Bratislava, 13--18 September 1993\\
 ~\\
 Sumbitted to {\sf ACTA PHYSICA SLOVACA}}

\maketitle

\begin{abstract}
Single-particle radiative mechanism of
$\gamma$ emission embedded into the pre-equ\-i\-li\-brium
exciton model is used to calculate the $\gamma$ emission
from a decay of $^{160}$Er$^{*}$ created in two
different ways. The initial stage of a reaction is
described using momentum-space overlaps of colliding
nuclei. We can reproduce the main features
of the observed $\gamma$ energy spectra including the
fact that the spectrum observed in more symmetric combination
is slightly harder than in the other case.
\end{abstract}

\section{Introduction}

A significant progress in the study of  $\gamma$ emission in heavy-ion
reactions can be marked within the last period, especially in collecting
the experimental data and application of various
bremsstrahlung mechanisms (see, e.g. \cite{Nife,Kama}).
Other approaches, namely the
quasideuteron model \cite{qd} suitable for intermediate energies,
and the exciton model of the $\gamma$ emission
\cite{BeDo,AG}, are less popular
in heavy-ion physics. However, the pre-equilibrium
exciton model has already firmly gained its ground
in reactions induced by nucleons
and light clusters at excitation energies of several tens of MeV,
and it starts to expand its applications towards
higher energies and especially heavier projectiles.

A great volume of available data on $\gamma$ emission from
heavy-ion reactions at $\gamma$ energies above 30 or 40
MeV has been
successfully analyzed in terms
of bremsstrahlung.
Emission of $\gamma$ quanta
below about 10 MeV (sometimes also at higher energies,
occasionally even
up to
the switch-on of the bremsstrahlung regime)
is usually attributed to
the equilibrium mechanism. A relatively overlooked energy
region from 10 to about 30 MeV is a potential
source of emerging inadequacies in description of $\gamma$
emission. Unfortunately, their manifestation is heavily
masked by other effects, so that the needs of more
precise description are thus suppressed. Practically,
the {\em exclusive} $\gamma$ spectra originated from a decay
of the same nucleus created by two different
projectile-target combinations, both leading to the
same composite system at practically the same excitation
energy and at very close angular momenta distributions,
are the only known type of data sensitive to possible refined
description of $\gamma$ emission mechanism. Such measurements
are unfortunately
rather seldom.

In our earlier paper \cite{Fizika}, we have been
interested in the decay of $^{132}$Nd$^{*}$
created once by $^{64}$Zn + $^{68}$Zn and by
$^{20}$Ne + $^{112}$Sn in the other case.
The {\em exclusive} $\gamma$ spectra from these two reactions
have been measured by Kamanin et al. \cite{Kama2}
and they demonstrate slight differences in the
$\gamma$ energy region from 8 to 15 MeV. This
has been explained by smaller number of degrees
of freedom in the early stage of the reaction
induced by lighter projectile, when compared to
the other case.

This time, we have been intrigued by the case of
$^{160}$Er$^{*}$ decay; when the composite system
originated from bombarding once by $^{16}$O
and the other time by $^{64}$Ni,
in both cases with the excitation energy $E=53$ MeV.
Though one would expect opposite, the observed {\em exclusive}
$\gamma$ spectra seem to be somewhat {\em harder} in the case of the
Ni-induced reaction here \cite{Thoe}.

\section{Exciton model $\gamma$ emission}

We employ the pre-equilibrium exciton model in our study.
Therein, the state of an excited nucleus is characterized
by its exciton number $n$ (particles above plus holes below
the Fermi level) and the excitation energy $E$. The reaction
proceeds from a relatively simple starting configuration,
characterized by the initial exciton number $n_0$. In the
course of a reaction, the nucleus develops via residual
interactions towards an equilibrium distribution.
In competition to the equilibration process, particle
as well as $\gamma$ emissions may occur.

For the energy region of 10 to 30 MeV, the single-particle
mechanism of the pre-equilibrium $\gamma$ emission
dominates.
Therein, just two processes
responsible for the $\gamma$ emission may occur,
associated with the exciton number change
$\Delta n = 0$ and $\Delta n = - 2$
\cite{BeDo,AG}. The corresponding $\gamma$ energy spectrum
can be expressed as
\begin{equation}
 \frac {{\mathrm d} \sigma} {{\mathrm d} \epsilon_\gamma} =
   \sigma_R \sum_n \tau_n \lambda_\gamma^c (n,E,\epsilon_\gamma)~,
\label{spektrum}
\end{equation}
with $\sigma_R$ denoting the reaction cross section
(that of a creation of a composite system) and $\tau_n$ being the
total time spent by a nucleus in the $n$-exciton state.
The $\gamma$ emission rates can be written as \cite{AG}
\begin{equation}
 \lambda_\gamma^c (n,E,\epsilon_\gamma) =
   \frac {{\epsilon_\gamma}^2 \sigma_a(\epsilon_\gamma)}
         {\pi^2 \hbar^3 c^2} ~
   \frac {\sum_{m=n,n-2} b(m,\epsilon_\gamma)
                         \omega(m,E-\epsilon_\gamma)}
         {\omega(n,E)}~~,
\label{garate}
\end{equation}
where $b$'s are the branching ratios for the two possible
processes, $\omega$'s are the exciton level densities,
and $\sigma_a(\epsilon_\gamma)$ is the
photo-absorption cross  section. Here, the experimental
data are preferred, but the Lorentzian shape (or a sum of
two Lorentzians for deformed nuclei) is frequently
used, either with global parameters or refering to the
existing tables of the GDR parameters
with individual widths, energies and peak
cross sections for various nuclei.

Obviously, one has to consider
the successive $\gamma$'s interspersed by nucleons
as needed to get
the total $\gamma$ production. The significantly enlarged set
of master equations of the exciton model,
which includes the coupling of different nuclei
and various excitation energies was employed to obtain
$\tau_n$'s.
The updated version of code PEQAG \cite{PEQAG}
has been used for the calculations.

\section{Initial configuration}

Till now we have introduced no specific features of the
heavy-ion processes. Indeed (if we are within the frame of the
pre-equilibrium exciton model), the main difference is contained
in the proper initial configuration of the reaction.
Commonly, one takes $n_0=1$ for nucleon- and $n_0=4$
for $\alpha$-induced reactions. In the collisions of
heavy ions, the same philosophy would lead to
$n_0=A_{proj}$, as was really used
in the first calculations of that kind \cite{blann}.
However, such a description has been found to
be somewhat inadequate, and the initial exciton number has been
a subject of separate studies.

As a first step, empirical systematics of $n_0/A_{proj}$ have
been reported \cite{CzJP,Korol}, which could serve as
a rough guideline. Consequently, a model has been developed,
which brings some insight into an understanding of
the dependence of the initial exciton number on the energy,
projectile mass and casually also the projectile-target
combination \cite{PRL,Fiz2,Ma}. The model is
based on calculating the overlap of colliding
partners in the momentum space.
According to \cite{Fiz2},
the calculated results at energies well above the Coulomb
barrier are grouped nearby line
\begin{equation}
  \frac {E}{n_{0}} = 4.6 + 0.54 \cdot \frac {E_{cm}-V_{C}}
     {A_{proj}}~,
\label{E/n0}
\end{equation}
whereas one should prefer
\begin{equation}
  \frac{n_0}{A_{proj}} = 0.09 + \left( 0.38 -
   0.08 \cdot \frac {A_{targ} - A_{proj}} {A_{targ} + A_{proj}} \right)
   \cdot \sqrt{ \frac {E_{cm}-V_C} {A_{proj}}}
\label{n0/Ap}
\end{equation}
as an approximate guide at smaller energies.
In the above equations, $E$ is the excitation energy of the
composite system, $E_{cm}$ the projectile c.m. energy, and
$V_C$ is the height of the Coulomb barrier.

\section{Results and discussion}

In our case (86.3 MeV $^{16}$O ions and 236.6 MeV $^{64}$Ni
beams), direct use of formula (\ref{n0/Ap}) yields
$n_0=6$ for O-induced and $n_0=11$ for Ni-induced reactions.
However, as we are very close to the Coulomb barrier in the
case of $^{64}$Ni beams, the influence of small additive
term on the r.h.s. of eq. (\ref{n0/Ap}), which has been
extracted with relatively large uncertainty, becomes
significant. Therefore a straightforward use
of (\ref{n0/Ap}) is not recommended and we have to apply
full overlap calculations as described in \cite{PRL}.
This procedure yields $n_0=7$ for the reaction induced
by $^{16}$O, but only $n_0=5$ for that by $^{64}$Ni.

The typical energy of the emitted particles at the
early stage of the reaction is
$\epsilon_{part} \approx E/n$, and that of $\gamma$'s
$\epsilon_{\gamma} \propto E/n$, so that the lower
exciton number implies higher mean energy of ejectiles,
and --- consequently --- harder spectra. We came to
the conclusion that the full account of the initial
stage of the reaction leads to {\em lower}
initial exciton number for
{\em more massive projectile} in our case, just the
opposite than one would expect! The experimental
data of exclusive $\gamma$ energy spectra have slopes close for
the two diffrent combinations, but the spectrum from
Ni-induced reaction is somewhat harder.
\begin{figure}
\begin{center}
    \mbox{\epsfig{file=gaspec.eps,height=8.5cm,width=12.0cm}}
      \caption{Measured
and calculated $\gamma$ energy spectra from
the decay of $^{160}$Er$^{*}$ created in two different
ways. Circles are
the experimental data of $^{16}$O + $^{144}$Nd
and crosses that of $^{64}$Ni + $^{96}$Zr
[8]; calculated spectra are drawn as lines:
dashed ($^{16}$O), thin ($^{64}$Ni with $n_0$ from
eq. (4)) and thick one ($^{64}$Ni with properly
calculated $n_0$).}
\end{center}
\end{figure}
Fig.~1
brings the comparison of experimental and calculated $
\gamma$ spectra.
Unfortunately, the available experimental data do not
extend
above 25 MeV. Rather fine differences at gamma energies 15 to
25 MeV do not enable to extract more than conditional
conclusions from a comparison to their theoretical
prediction.
The calculated spectra are relatively close to
the measured ones, and the $\gamma$ spectrum from Ni-induced
reaction calculated with proper tracing the initial stage
of the reaction in the overlaps of the colliding nuclei
is harder than that from O-induced case, just in accord
with the data.
 In neither of the cases the pre-equilibrium
 description fully reproduces the data, and some
 minor differences between
 the calculations and the experiment remain in both
 projectile-target combinations considered;
 e.g., the
 calculated spectra are somewhat below the experimental ones
 at energies exceeding 16 MeV.
The main and striking feature of the data, namely
that the $\gamma$'s from more symmetric combination
are harder than in the asymmetric case, is caught and
reproduced in our calculations; whereas
other approaches fail to reproduce this specific feature
of the data.

\section{Conclusions}

We are aware of the fact that we have not arrived to
a sufficient and
satisfactory agreement of the calculations to the
data. The pre-equilibrium exciton model including all
its ingredients (level densities, initial exciton
number, transition matrix element, etc.) is very schematic
and rough, and one cannot expect a complete fit here.
Our calculations have been performed just with
global parameters both for the model and for the
$\gamma$ emission itself, so that they lead to an
{\em a priori} result.

By sophisticated adjustment of the parameters, one will
be able to improve the quality of the fit, but such
a calculation will not bring more physical information
than the present one due to
the underlying shortcomings of the model.

Finally, we conclude that we are able to catch the main
features of the observed $\gamma$ spectra
below 25 MeV, namely their slope and its possible
variation with changing projectile-target combination
for heavy-ion collisions at moderate energies within a
simple use of the single-particle radiation mechanism
in connection with the pre-equilibrium
exciton model. This includes also the
experimental aspects which are not reproducible by
other approaches.

\vspace*{1cm}
\noindent
The work has been supported
in part by Slovak Academy of Sciences Grant No. 042/93
and by International Atomic Energy Agency Contract No. 7811/RB.


\vspace{1cm}

\begin{thebibliography}{99}

\bibitem{Nife} H. Nifenecker and J.A. Pinston, Prog. Part. Nucl. Phys.
               {\bf 23} (1989), 271;
               {\em and} Annu. Rev. Nucl. Part. Sci. {\bf 40}
               (1990), 113.
\bibitem{Kama} V.V. Kamanin {\em et al.}, Particles and Nuclei {\bf 20}
               (1989), 741.
\bibitem{qd}   P. Oblo\v{z}insk\'{y}, Phys. Rev. {\bf C40} (1989), 1591.
\bibitem{BeDo} E. B\v{e}t\'{a}k and J. Dobe\v{s}: Phys. Lett.
               {\bf 84B}, 368 (1979)
\bibitem{AG} J.M. Akkermans and H. Gruppelaar: Phys. Lett.
               {\bf 157B}, 95 (1985)
\bibitem{Fizika} E. B\v{e}t\'{a}k, Fizika {\bf 19} (1987), Suppl. 1, 49.
\bibitem{Kama2} V.V. Kamanin et al., Z. Phys. A327 (1987), 109
               {\em and in} Internat. School--Seminar Heavy-Ion Phys.,
               Dubna 1986 (D7-87-68, Dubna 1987), p. 489
\bibitem{Thoe} M. Thoennessen {\em et al.}, Talk presented at
               Workshop Interface between Nucl. Struct. and
               Heavy-Ion Reaction Dynamics, Notre Dame May
               1990 (Report ORNL, {\em (unnumberred)});\\
               M. Thoennessen and J.R. Beene, Talk at Symp. Reflect.
               Direct. in Low Energy Heavy Ion Phys., Oak Ridge
               October 1991 (Report MSUCL-799)
               {\em (Proceedings in press)}
\bibitem{PEQAG} E. B\v et\'ak, Report INDC(CSR)-016/LJ
               IAEA Vienna 1989 {\em (code available
               on request)}
\bibitem{blann} M. Blann, Nucl. Phys. {\bf A235} (1974), 211.
\bibitem{CzJP} E. B\v{e}t\'{a}k, Cz. J. Phys. {\bf B34} (1984), 850;
\bibitem{Korol} M. Korolija {\em et al.}, Phys. Rev. Lett. {\bf 60}
               (1988), 193.
\bibitem{PRL} N. Cindro {\em et al.}, Phys. Rev. Lett. {\bf 66} (1991),
               868.
\bibitem{Fiz2} N. Cindro et al., Fizika {\bf B1} (1992), 51
\bibitem{Ma}   Y.G. Ma et al., Phys. Rev. {\bf C48} (1993), 448

\end{thebibliography}
\end{document}